\documentclass[prd,amssymb,amsmath,eqsecnum,nofootinbib,preprintnumbers]{revtex4-1}
\pdfoutput=1
\usepackage{verbatim,slashed,graphics,graphicx,color,slashed,textcomp,bm,diagbox,tabu}
\usepackage[utf8]{inputenc}

\graphicspath{{figures/}}

\usepackage[colorlinks=true,linkcolor=red,urlcolor=blue,citecolor=blue]{hyperref}

\newcommand{\GeV}{\mathrm{GeV}}
\newcommand{\TeV}{\mathrm{TeV}}

\begin{document}
	
\title{On Thermal Gravitational Contribution to Particle Production and Dark Matter}
\preprint{UT-17-27}

%\begin{comment}

\author{Yong Tang$^{a}$ and Yue-Liang Wu$^{b,c,d}$}
\affiliation{
${}^a$Department of Physics, Faculty of Science, \\
The University of Tokyo, Bunkyo-ku, Tokyo 113-0033, Japan\\
${}^b$International Centre for Theoretical Physics Asia-Pacific, Beijing, China \\
${}^c$Institute of Theoretical Physics, Chinese Academy of Sciences, Beijing 100190, China\\
${}^d$University of Chinese Academy of Sciences, Beijing 100049, China}

\begin{abstract}
We investigate the particle production from thermal gravitational annihilation in the very early universe, which is an important contribution for particles that might not be in thermal equilibrium or/and might only have gravitational interaction, such as dark matter (DM). For particles with spin $0, 1/2$ and $1$ we calculate the relevant cross sections through gravitational annihilation and give the analytic formulas with full mass-dependent terms. We find that DM with mass between $\TeV$ and $10^{16}\GeV$ could have the relic abundance that fits the observation, with small dependence on its spin. We also discuss the effects of gravitational annihilation from inflatons. Interestingly, contributions from inflatons could be dominant and have the same power dependence on Hubble parameter of inflation as that from vacuum fluctuation. Also, fermion production from inflaton, in comparison to boson, is suppressed by its mass due to helicity selection. 

\end{abstract}

 \maketitle 
 
%\end{comment}

%\subsection*{On Thermal Gravitational Contribution to Particle Production and Dark Matter} 

\section{Introduction}\label{sec:intro}
The accumulated firm evidence for dark matter (DM) has challenged modern particle physics for decades. From the galactic rotation curves to galaxy cluster, large scale structure (LSS) and cosmic microwave background (CMB), the existence of DM has been well-established, based only on the gravitational interaction. Numerous models for DM has also been proposed, see Refs.~\cite{Bertone:2004pz, Feng:2010gw} for reviews. Broadly speaking, for DM as elementary particles, it either can be  in thermal equilibrium with other particles and then freeze out, or was never in equilibrium but still produced gradually through various processes. The first class is usually referred as weakly-interacting massive particle (WIMP), while the second includes axion, sterile neutrino, gravitino and so on. 

All the mentioned DM candidates above inevitably have interactions other than gravitation, therefore in principle could give rise to possible signatures in direct, indirect and collider searches. However, so far there is no confirmed evidence in all those searches for DM's new interaction, it is fair to ask what if DM only has gravitational interaction. Recent studies~\cite{Garny:2015sjg, Tang:2016vch} have shown that it is viable to generate scalar DM abundantly with only gravitational annihilation, namely particles in the thermal both annihilate into DM through a graviton, the quantum of Einstein's gravity in the weak-field limit. Scenarios and phenomenologies in extended theories are also discussed, for example in Refs.~\cite{Ren:2014mta, Cata:2016dsg, Cata:2016epa, Nagoya:2016yir, Babichev:2016hir, Babichev:2016bxi}. 

In the view of effective field theories, microscopically gravity can be treated effectively as quantum field theory as long as the energy scale is much lower than Planck scale ($M_P=1.12\times 10^{19}\GeV$)~\cite{Donoghue:1994dn, Burgess:2003jk}. Recently, it has also been shown that general relativity can be derived as an effective field theory of gravitational quantum field theory with spin and scaling gauge symmetries~\cite{Wu:2015wwa, Wu:2017rmd}. Since the energy scale during/after inflation has already been constrained to be $\lesssim 10^{16}\GeV$ which is much lower than $M_P$, we would expect that the local scattering and/or annihilation through graviton can be described in effective field theory. Then these processes should in principle contribute to the cosmological evolution of all particle species, including DM. Particles with interactions much stronger than gravity would be in thermal equilibrium with other particles and short-range gravitational processes are essentially irrelevant for them. However, if DM is very weakly interacting and was never in equilibrium in the early universe, we should include the contributions from gravitational processes. 

In this paper, we investigate the viable mass range for DM with spin $0, 1/2$ and $1$, produced by the gravitational annihilation of particles in the thermal bath with various spins. We compute all the possible, general annihilation cross sections analytically, including all the finite mass term. We find that for the production from particles in the thermal bath the abundance of DM is tightly related with the highest temperature $T_{\mathrm{max}}$ after inflation, proportional to $T^3_{\mathrm{max}}/M^3_P$ if its mass $m_X<T_{\mathrm{max}}$ and $m^3_X/M^3_P \exp{[-2m_X/T_{\mathrm{max}}]}$ if  $m_X>T_{\mathrm{max}}$. We also discuss the effects from inflation dynamics and show that, gravitational annihilation from inflatons might be the dominant channel for scalar/vector DM production (there is a suppression factor for fermionic DM due to helicity selection) and interestingly has the same power dependence on Hubble parameter as production from vacuum fluctuation. 

This paper is organized as follows. In Sec.~\ref{sec:intro} we start with the standard Boltzmann equation to follow the cosmological evolution of particles and establish the convention and terminology for later discussions. Then in Sec.~\ref{sec:cs} we calculate the gravitational annihilation cross section for different initial and final states with spin 0, 1/2 and 1. Later in Sec.~\ref{sec:DM} we apply our calculated cross section to DM and investigate the viable mass range. In Sec.~\ref{sec:inflation} we discuss the effects from chaotic inflation and show that inflaton's contribution can be very important. Finally, we give the summary.

\section{Boltzmann Equation}
To be self-contained, let us start with the standard Boltzmann equation in cosmology~\cite{Kolb:1990vq} for the evolution of number density $n_3$ through the $2\leftrightarrow 2$ process\footnote{Following the same formalism, processes with multiple initial or final states can also be included. These contributions could also be important unless they are suppressed by additional small couplings or phase space factors.}, $p_1+p_2\leftrightarrow p_3+p_4$,
\begin{align}
\dot{n}_3+3Hn_3\equiv \frac{d\left(a^3n_3\right)}{a^3dt}=&\int \frac{d^3\bm{p}_1}{(2\pi)^3 2E_1}\frac{d^3\bm{p}_2}{(2\pi)^3 2E_2}\frac{d^3\bm{p}_3}{(2\pi)^3 2E_3}\frac{d^3\bm{p}_4}{(2\pi)^3 2E_4}(2\pi)^4\delta^4(p_1+p_2-p_3-p_4) \times \nonumber \\
& \sum_{\mathrm{pol}} \left[ f_1f_2(1\pm f_3)(1\pm f_4)\left| \mathcal{M}_{12\rightarrow 34}\right|^2 - f_3f_4(1\pm f_1)(1\pm f_2)\left| \mathcal{M}_{34\rightarrow 12}\right|^2  \right],\label{eq:boltz}
\end{align}
where $a$ is the scalar factor, Hubble parameter $H=\dot{a}/a$, $\bm{p}_i$ denote the spatial momenta, $p_i$ for 4-vector, $\mathcal{M}$ is the matrix element, $f_i$ is the distribution for particle $i$ without internal degree of freedom, $+ (-)$ sign in $\pm$ is for bosons (fermions) and $\sum_{\mathrm{pol}}$ means the sum of all polarizations. For particles that were in thermal equilibrium, such as WIMP, we need to keep both terms in the bracket of Eq.~(\ref{eq:boltz}). This is due to the cross symmetry $\mathcal{M}_{12\rightarrow 34}=\mathcal{M}_{34\rightarrow 12}$ and $f_1f_2$ is compatible to $f_3f_4$ for $E_i\sim m_3$ where $m_3$ is the mass for particle $3$. In cases where $f_{3,4}$ is much smaller than $1$ and/or $f_{1,2}$, we can neglect the second term and the above Boltzmann equation becomes
\begin{align}
\frac{d\left(a^3n_3\right)}{a^3dt}=&\int\frac{f_1d^3\bm{p}_1}{(2\pi)^3 2E_1}\frac{ f_2d^3\bm{p}_2}{(2\pi)^3 2E_2}\left[\frac{d^3\bm{p}_3}{(2\pi)^3 2E_3}\frac{d^3\bm{p}_4}{(2\pi)^3 2E_4} (2\pi)^4\delta^4(p_1+p_2-p_3-p_4)  \sum_{\mathrm{pol}} \left| \mathcal{M}_{12\rightarrow 34}\right|^2 \right] , 
\end{align}
The term in the bracket can be replaced by $4 \mathcal{F} g_1 g_2 \sigma_{12\rightarrow 34}$, where $g_i$ is the spin degree of freedom, $\sigma\equiv \sigma_{12\rightarrow 34}$ is the cross section and $\mathcal{F}=[(p_1\cdot p_2)^2-m_1^2m_2^2]^{1/2}$. So we have
\begin{align}
\frac{d\left(a^3n_3\right)}{a^3dt}=&\int\frac{f_1 g_1 d^3\bm{p}_1}{(2\pi)^3 E_1}\frac{ f_2 g_2 d^3\bm{p}_2}{(2\pi)^3 E_2} F   \sigma, \label{eq:master}
\end{align}
Changing to the integration variables $E_1,E_2$ and $s$, we have
\begin{equation}
d^3\bm{p}_1d^3\bm{p}_2 = 4\pi^2 E_1E_2 d E_1 dE_2 ds = 2\pi^2 E_1E_2 d E_+ dE_- ds,
\end{equation}
where $E_+=E_1+E_2,E_-=E_1-E_2$, and $s=(p_1+p_2)^2$. As will be shown in next section, throughout our discussion, we have $m_1=m_2=m$ and $m_3=m_4=M$ and the integration range then can be simplified to 
\begin{equation}
s \geq \mathrm{max}(4m^2,4M^2) , E_1 \geq m, E_2 \geq m, E_+\geq \sqrt{s}, \left|E_-\right| \leq \sqrt{ 1-4m^2/s }\sqrt{E^2_+-s}.
\end{equation} 

So far, the discussions have been quite general and apply for other very weakly interacting particles as well, see Ref.~\cite{Bernal:2017kxu} for a recent review. It is evident that the key part is to calculate the annihilation cross section $\sigma$. After that we can perform numerical integration or analytic computation for some special cases. If $f_{1,2}$ have quantum statistical distributions, like Fermi-Dirac or Bose-Einstein distributions $(e^{E/T}\pm 1)^{-1}$, no compact analytic formulas can be derived. However, for $E > T$, we can use approximate Maxwell-Boltzmann distribution, $e^{-E/T}$, and then integrate over $E_-$ and $E_+$ to get
\begin{equation}
\frac{d\left(a^3n_3\right)}{a^3dt} = \frac{g^2_1 T}{32\pi^4}\int ds \,\sigma\, \sqrt{s}(s-4m^2)K_1\left(\frac{\sqrt{s}}{T}\right),
\end{equation}
where $K_i$ is the modified Bessel function of the second kind with order $i$. 

\section{Annihilation Cross Section}\label{sec:cs}

\begin{figure}[t]
	\includegraphics[width=0.3\textwidth,height=0.18\textwidth]{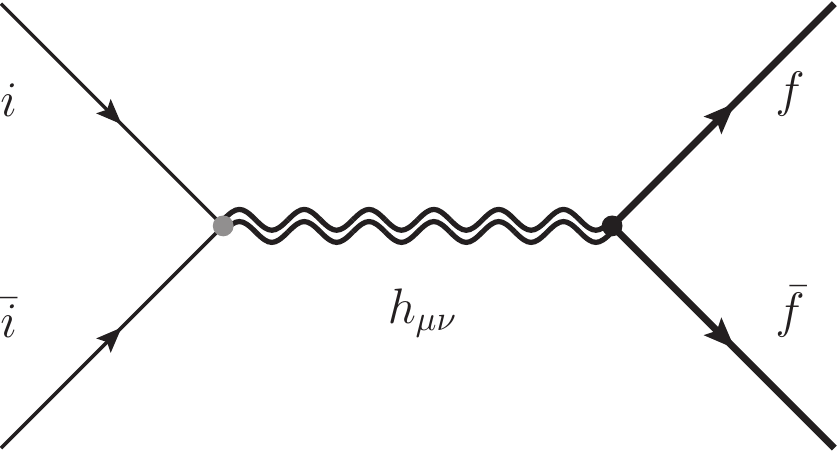}
	\caption{Annihilation process for $i\rightarrow f$, where particles $i$ and $f$ can be scalars $S$, fermions $F$ (spin 1/2), massive vectors $V$ and massless vectors $\gamma$. For massive particles, we always denote the initial states'mass as $m$ and the final states' as $M$. The double lines represent the graviton field, $h_{\mu\nu}$. Arrows mean the directions of momenta. Note that although $i$ and $f$ might have the same spin, they have to be different particles to affect the number density in Boltzmann equation.   \label{fig:anni}}
\end{figure}

In this section, we compute the annihilation cross section in the center-of-mass (CM) frame for various initial and final states in Fig.~\ref{fig:anni}. Note that the initial particles are different from the final ones so that the process can change the number density and contribute to Boltzmann equation, although in a broader context for other physics problems they can be the same. Since the cross section is a Lorentz-invariant quantity, the results derived here will also be valid in other frames.

In effective field theory, the leading interactions between graviton and matter are described by 
\begin{equation}
\mathcal{L}_{\mathrm{int}}=\frac{\kappa}{2}h_{\mu\nu}T^{\mu\nu},
\end{equation}
where $\kappa=\sqrt{32\pi G}$($G$ is the Newton's constant), $h_{\mu\nu}$ is the graviton field and $T_{\mu\nu}$ is the energy-momentum tensor for matter fields. This term is linear on $h_{\mu\nu}$ but sufficient for our discussions in which less than $2$ gravitons appear in the processes. We shall use the harmonic gauge fixing condition for gravity so that the graviton's propagator with momentum $p$ has the following form
\begin{equation}
G_{\mu\nu;\rho\sigma}(p)=\frac{i}{2p^2}\left(\eta_{\mu\rho}\eta_{\nu\sigma}+\eta_{\mu\sigma}\eta_{\nu\rho} - \eta_{\mu \nu}\eta_{\rho\sigma} \right)
\end{equation} 
where $\eta_{\mu\nu}$ is the metric for flat spacetime. Since we are considering the leading-order tree-level scattering processes, we do not need to include higher-dimensional operators with more graviton $h_{\mu\nu}$, renormalization effects and ghost. 

The symmetric energy-momentum tensors $T^{\mu\nu}$ for complex scalar $S$, spin-$\frac{1}{2}$ Dirac fermion $F$, massive vector $V$ and massless vector $\gamma$ are listed in the following
\begin{align} 
T^{\mu\nu}_S=&-\eta^{\mu\nu}\partial ^\alpha S^\dagger \partial _\alpha S + \eta^{\mu\nu}m_S^2S^\dagger S + \partial^\mu S^\dagger \partial ^\nu S +\partial^\nu S^\dagger \partial ^\mu S, \\
T^{\mu\nu}_F=&-\eta^{\mu\nu}\left(\overline{F}i\slashed{\partial}F-m_F\overline{F}F\right)+ \frac{1}{2}\overline{F}i\gamma^\mu \partial^\nu F + \frac{1}{2}\overline{F}i\gamma^\nu \partial^\mu F \nonumber \\
&  + \frac{1}{2}\eta^{\mu \nu} \partial^\alpha \left(\overline{F}i\gamma_\alpha F\right)-\frac{1}{4}\partial^\mu \left(\overline{F}i\gamma^\nu F \right) -\frac{1}{4}\partial^\nu \left(\overline{F}i\gamma^\mu F \right), \\
T^{\mu\nu}_V=&\eta^{\mu\nu}\left(\frac{1}{4}F^{\alpha\beta}F_{\alpha\beta}-\frac{1}{2}m_V^2V^\alpha V_\alpha \right) 
- \left(F^{\mu\alpha}F^{\nu}{}_\alpha - m_V^2 V^\mu V^\nu \right),\\
T^{\mu\nu}_\gamma =&\frac{1}{4}\eta^{\mu\nu} F^{\alpha\beta}F_{\alpha\beta} - F^{\mu\alpha}F^{\nu}{}_\alpha.
\end{align}
$T^{\mu\nu}$ for real scalar $\phi$ can easily be obtained by substituting $S=(\phi +i\varphi)/\sqrt{2}$. For scalar with nonminimal coupling $\zeta S^\dagger S R$ we should add $2\zeta(\partial^\mu \partial^\nu - \eta^{\mu\nu}\partial_\alpha \partial^\alpha ) S^\dagger S$.  Then we can get the Feynman rules to do the calculation of Fig.~\ref{fig:anni}. To make the results as compact as possible, we extract the common factor for unpolarized collisions $p_i+p_{\bar{i}}\rightarrow p_f+p_{\bar{f}}$,
\begin{equation}
\sigma=\frac{1}{32\pi s \left(S  g^2_i\right) } \frac{\left|\vec{p}_{f}\right|}{\left|\vec{p}_{i}\right|}\int d\cos\theta{\sum_{\mathrm{pol}}}\left|\mathcal{M}\right|^{2}
\equiv \frac{\kappa^4}{32\pi s \left(S g^2_i\right) } \frac{\left|\vec{p}_{f}\right|}{\left|\vec{p}_{i}\right|} \mathcal{A} , \label{eq:cross}
\end{equation}
where $g_i$ is the degrees of freedom for initial state $i$, $S$ is the symmetric factor ($S=2$ for identical final states, for example, real scalars, neutral gauge bosons, otherwise $S=1$), $\left|\vec{p}_{i}\right|$ and $\left|\vec{p}_{f}\right|$ are the lengths of three-momentum for initial and final states, respectively. As shown, $\mathcal{M}$ is the matrix element and we have defined $\mathcal{A}$ as the integration of polarization-summed $\sum \left|\mathcal{M}\right|^{2}$ over the scattering angle $\theta$, with the $\kappa^4$ factor pulled out. 

Note that the kinematic variables in CM frame for $m_i=m_{\bar{i}}\equiv m$ and $m_f=m_{\bar{f}}\equiv M$, and
\begin{align}
\left|\vec{p}_{i}\right|=\sqrt{E_{i}^{2}-m^{2}}, \; \left|\vec{p}_{f}\right|=\sqrt{E_{f}^{2}-M^{2}},  E_{i,f}=\sqrt{s}/2.
\end{align}
After some tedious calculations, we obtain $\mathcal{A}$ for different processes of initial states with mass $m$ and final ones with $M$ where both the initial and final states can be complex scalar $S$, fermion $F$ (spin 1/2), massive vector $V$ and massless vector $\gamma$. For processes involving final scalar $S$, 
\begin{align}
\mathcal{A}\left(S\rightarrow S\right)=&\frac{7 m^4 M^4}{30s^2}-\frac{ m^2 M^2 }{30s}\left(m^2+M^2\right), \nonumber\\ 
&+\frac{1}{40} \left(m^4+4 m^2 M^2+M^4\right) +\frac{s}{120} \left(m^2+M^2\right)+\frac{s^2}{240},\\
\mathcal{A}\left(F\rightarrow S\right)=&-\frac{7 m^4 M^4}{15s^2}-\frac{ m^2 M^2 }{60s} (M^2-4 m^2) \nonumber \\ 
&+\frac{1}{60} \left(2 M^4+3 m^2 M^2-3 m^4\right) -\frac{s}{240}  (4M^2-m^2)+\frac{s^2}{480},\\
\mathcal{A}\left(V\rightarrow S\right)=&\frac{101 m^4 M^4}{30s^2}-\frac{m^2 M^2}{10s}\left(11 M^2+m^2\right) \nonumber \\ 
& + \frac{1}{120}\left(19 M^4+76 m^2 M^2+49 m^4\right)-\frac{7s}{120}\left(m^2+M^2\right)+\frac{s^2}{80},\\
\mathcal{A}\left(\gamma\rightarrow S \right)=&\frac{1}{120}\left(s-4 M^2\right)^2,
\end{align}
for fermion
\begin{align}
\mathcal{A}\left(F\rightarrow F\right)=&\frac{14 m^4 M^4}{15 s^2}+\frac{ m^2 M^2 }{30s}\left(m^2+M^2\right), \nonumber \\ 
&-\frac{1}{120}\left(8 m^4-3 m^2 M^2+8 M^4\right)-\frac{s}{120} \left(m^2+M^2\right)+\frac{s^2}{160},\\
\mathcal{A}\left(V\rightarrow F\right)=&-\frac{101 m^4 M^4}{15s^2}+\frac{ m^2 M^2 }{20s}\left(44 M^2-m^2\right)\nonumber \\ 
&-\frac{1}{60}\left(19 M^4-19 m^2 M^2-26 m^4\right) -\frac{s}{240} \left(7 M^2+52 m^2\right)+\frac{13 s^2}{480},\\
\mathcal{A}\left(\gamma\rightarrow F\right)=&\frac{1}{120}\left(s-4 M^2\right)(3s+8M^2),
\end{align}
and for vector
\begin{align}
\mathcal{A}\left(V\rightarrow V\right)=&\frac{2983 m^4 M^4}{30s^2}-\frac{293 m^2 M^2 }{10s}\left(m^2+M^2\right),\nonumber \\ 
&+\frac{1}{120} \left(257 m^4+1188 m^2 M^2+257 M^4\right)-\frac{37 s}{40} \left(m^2+M^2\right)+\frac{29 s^2}{240},\\
\mathcal{A}\left(\gamma\rightarrow V \right)=&\frac{13}{120}\left(s-4 M^2\right)^2,\\
\mathcal{A}\left(\gamma \rightarrow \gamma \right)=&\frac{s^2}{10}.
\end{align}
Note that we can use the cross symmetry, $\mathcal{A}\left(f\rightarrow i \right)=\mathcal{A}\left(i\rightarrow f \right)$ with interchanging $m\leftrightarrow M$, to get $\mathcal{A}$s for other processes, such as $\mathcal{A}\left(S\rightarrow F\right)$, $\mathcal{A}\left(S/F\rightarrow V\right)$ and $\mathcal{A}\left( S/F/V \rightarrow \gamma \right)$. In the case $s\gg 4m^2$ and $s\gg 4M^2$, we can neglect the mass-dependent terms and get very concise $\mathcal{A}$s which are just proportional to $s^2$.

The above results have shown consistencies under several checks. For example, $\mathcal{A}$ is gauge invariant when involving massless vector $\gamma$ where we have explicitly checked in $R_\xi$ gauge and the results are independent of gauge-fixing parameter $\xi$ in the $T^{\mu\nu}_\gamma(\xi)$,
\begin{align} 
T^{\mu\nu}_\gamma \left(\xi\right)=&\frac{1}{4}\eta^{\mu\nu} F^{\alpha\beta}F_{\alpha\beta} - F^{\mu\alpha}F^{\nu}{}_\alpha -\frac{1}{\xi}\eta^{\mu\nu} \left[\partial^\alpha \partial ^\beta \gamma_\alpha \gamma_\beta - \frac{1}{2}\left(\partial^\alpha \gamma_\alpha\right)^2 \right] +\frac{1}{\xi}\left(\partial^\mu \partial^\alpha \gamma_\alpha \gamma^\nu + \partial^\nu \partial^\alpha \gamma_\alpha \gamma^\mu \right).
\end{align}
And the coefficient of $s^2$ term in $\mathcal{A}\left(V\rightarrow S \right)$ is three times as that in $\mathcal{A}\left(S\rightarrow S \right)$, which is due to three polarizations of $V$. Furthermore, $\mathcal{A}\left(i\rightarrow f \right)$s are symmetric over $m$ and $M$ when the initial and final states are the same, $i=f$. For later convenience, we also tabulate the case $m^2\ll s$ in Table.~\ref{tab:As}. One can easily check that $\mathcal{A}\left(V\rightarrow f \right)=\mathcal{A}\left(S\rightarrow f \right)+\mathcal{A}\left(\gamma\rightarrow f \right)$. Interestingly, we notice that $\mathcal{A}\left(\gamma\rightarrow f \right)=4\mathcal{A}\left(F\rightarrow f \right)$ which might be related with spin structures in gravitational interactions. 

\section{Application to Dark Matter}\label{sec:DM}
\begin{table}[bt]
	\begin{tabular}{|c|c|c|c|c|}
		\hline 
		\diagbox{Initial $i(m)$}{Final $f(M)$} &  $S$  & $F$ & $V$ & $\gamma$	\\ [5pt]
		\hline 
		$S$ & $\dfrac{ M^4}{40}+\dfrac{s M^2}{120}+\dfrac{s^2}{240}$ & $\dfrac{1}{480}\left(s-4 M^2\right)(s+6M^2) $ & $\dfrac{49M^4}{120}-\dfrac{7sM^2}{120}+\dfrac{s^2}{80} $ & $\dfrac{s^2}{120}$\\ [5pt]
		\hline
		$F$ & $\dfrac{1}{480}\left(s-4M^2\right)^2$ &$ \dfrac{1}{480}\left(s-4 M^2\right)(3s+8M^2)$ & $\dfrac{13}{480}\left(s-4 M^2\right)^2 $ & $\dfrac{s^2}{40}$ \\ [5pt]
		\hline
		$V$ & $\dfrac{19 M^4}{120}-\dfrac{7sM^2}{120}+\dfrac{s^2}{80}$ & $\dfrac{1}{480}\left(s-4 M^2\right)(13s+38M^2)$ & $\dfrac{257 M^4}{120} -\dfrac{37 sM^2 }{40} + \dfrac{29 s^2}{240}$ & $\dfrac{13s^2}{120} $ \\ [5pt]
		\hline
		$\gamma$&$\dfrac{1}{120}\left(s-4 M^2\right)^2$ & $\dfrac{1}{120}\left(s-4 M^2\right)(3s+8M^2)$ & $\dfrac{13}{120}\left(s-4 M^2\right)^2$ & $\dfrac{s^2}{10}$ \\ [5pt]
		\hline 
	\end{tabular}
	\caption{$\mathcal{A}$ for the case $m^2/s\rightarrow 0 $ (initial states with mass $m$ and final states with mass $M$). Note that the results are not symmetric under $i\leftrightarrow f$ since we do not take the limit $M^2/s\rightarrow 0 $. \label{tab:As}}
\end{table}

With the cross section in hand, we now proceed to compute the abundance for stable particles like DM $X$ with mass $M=m_X$. In the absence of entropy production, we have $d(a^3 \bm{s})/dt=0$, where $\bm{s}$ is the entropy density. Therefore, we can rewrite the equation Eq.~\ref{eq:master} in terms of a more convenient quantity, th yield $Y_X\equiv n_X/\bm{s}$,  
\begin{equation}
\frac{dY_X}{d t} = \frac{g^2_1 T}{32\pi^4}\int ds \,\sigma\, \sqrt{s}(s-4m^2)K_1\left(\frac{\sqrt{s}}{T}\right).
\end{equation}
In the radiation dominant era, we have the following relations,
\[
H^2=\frac{8\pi G \rho_r}{3}\equiv\frac{\kappa^2\rho_r}{12},\,\rho_r= \dfrac{\pi^2}{30}g_{\ast} T^4,\;dt=-\frac{dT}{HT},
\]
where $g_{\ast}$ is the total number of effectively massless degrees of freedom. Integrate over temperature from the minimal value to maximum one, we finally get 
\begin{equation}
Y_X=\int_{T_{\mathrm{min}}}^{T_{\mathrm{max}}}\frac{dT}{HT\bm{s}} \left[\frac{g^2_1 T}{32\pi^4}\int ds \,\sigma\, \sqrt{s}(s-4m^2)K_1\left(\frac{\sqrt{s}}{T}\right) \right].\label{eq:yield}
\end{equation}
The above result has negligible dependence on $T_\mathrm{min}$, so we can freely take $T_\mathrm{min}$ as zero or the present temperature of CMB. The yield $Y_X$ is related with the observed energy fraction for DM $\Omega_X$ at present time,
\begin{equation}
\Omega_X = \frac{\Omega_b m_X }{m_p n_\gamma \eta }\bm{s}_0 Y_X, 
\end{equation}
where $\Omega_b$ is the energy density fractions of baryon, $m_p\simeq 1\GeV$ is proton mass, $n_\gamma$ is the number density of photon today, $\bm{s}_0$ is the total entropy density of photon and neutrino, and $\eta\simeq 6\times 10^{-10}$ is baryon-to-photon ratio. Assuming a minimal particle content in thermal bath, namely only SM, and the temperature is higher than electroweak symmetry breaking, all SM particles are therefore massless and we can use the results in Table.~\ref{tab:As}. Since our formalism is for Dirac fermions, there is a factor of $1/2$ for Weyl particles (neutrino in SM). Taking all these into account, we have $2$ complex scalars, $45/2 (1/2\times 3 + 1\times3 + 2\times 3\times3) $ Dirac fermions and $12 (8+3+1)$ massless gauge bosons.

\begin{figure}[t]
	\includegraphics[width=0.56\textwidth,height=0.46\textwidth]{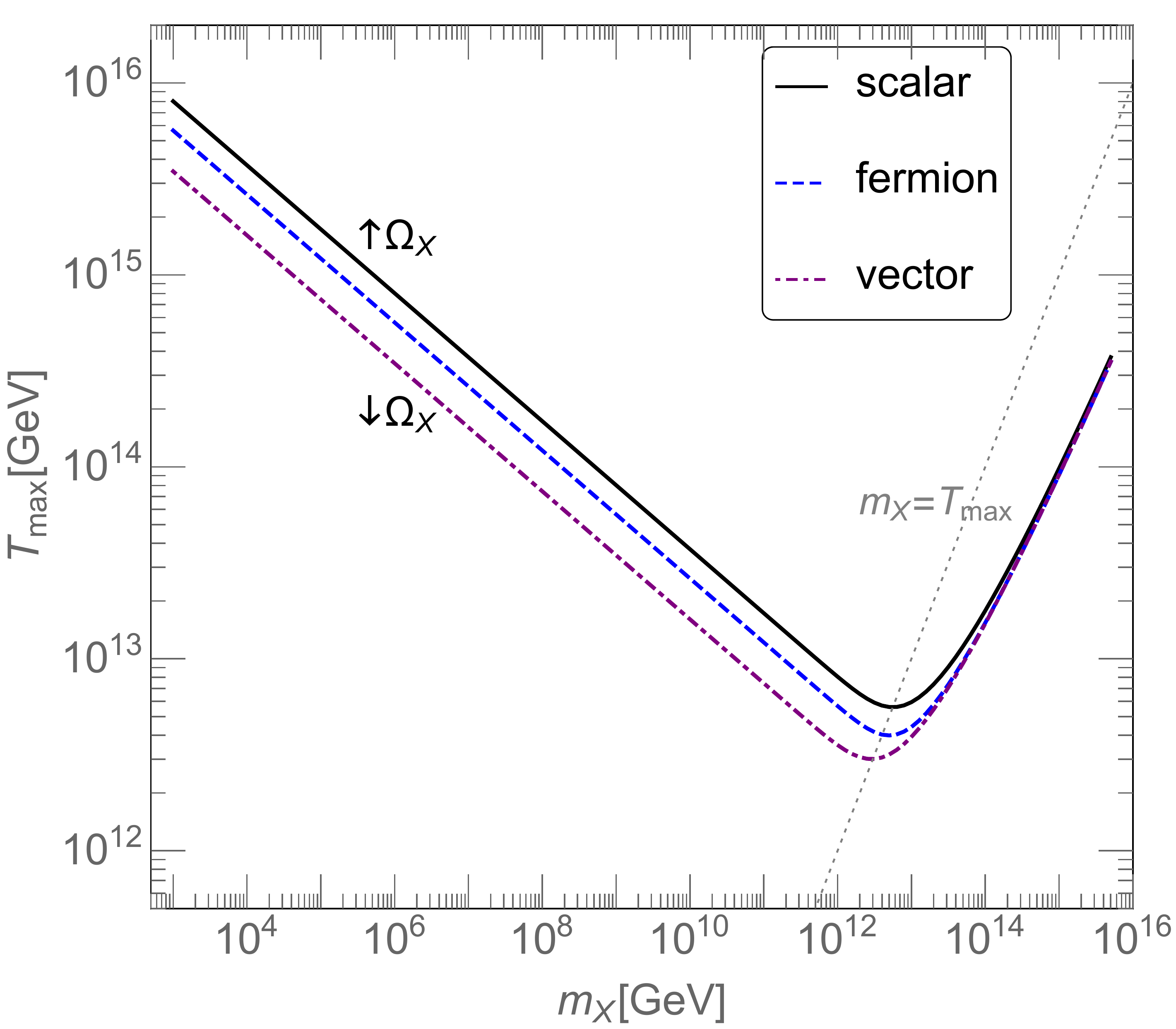}
	\caption{Numeric illustration of the correlation between temperature $T_\mathrm{max}$ and DM mass $m_X$, in spin-0 (solid black line), spin-$\frac{1}{2}$ (blue dashed line) and spin-1 (purple dot-dashed) cases. All these curves indicates that $\Omega_X\simeq 0.258$. Going to region above (below) the curve would increase (decrease) $\Omega_X$. The gray dotted line marks $m_X=T_{\mathrm{max}}$ and on its left (right) side $m_X<T_{\mathrm{max}}$ ($m_X>T_{\mathrm{max}}$).    \label{fig:relic}}
\end{figure}

In Fig.~\ref{fig:relic}, we illustrate the correlation between temperature $T_\mathrm{max}$ and DM mass $m_X$ for fixed $\Omega_X\simeq 0.258$~\cite{Planck:2015xua} in several cases, by integrating Eq.~(\ref{eq:yield}) numerically. DM with spin zero, 1/2 and 1 are shown with solid, dashed and dot-dashed curves, respectively. The gray dotted line indicates $m_X=T_{\mathrm{max}}$, while its left (right) side marks $m_X<T_\mathrm{max}$ ($m_X>T_\mathrm{max}$). The turnover of these curves at $m_X=T_{\mathrm{max}}$ is due to the following reasons. When $m_X<T_\mathrm{max}$, $\Omega_X$ is proportional to $\kappa^3T_{\mathrm{max}}^3$ and increasing $m_X$ would require smaller $T_{\mathrm{max}}$ for fixed $\Omega_X$, which is exactly the reason why we see straight lines in the logarithmic plot. This temperature dependence is different from gravitino production~\cite{Moroi:1993mb} that depends on $T_{\mathrm{max}}$ linearly because gravitino is mostly produced by the scattering of supersymmetric particles, not by the diagram in Fig.~\ref{fig:anni}. For $m_X>T_\mathrm{max}$, the production is exponentially suppressed due to the Boltzmann distribution at high energy, roughly scaling as $\kappa^3m_X^3\exp{[-2m_X/T_{\mathrm{max}}]}$. In such a circumstance, $\Omega_X=0.258$ would require $T_{\mathrm{max}}\sim m_X/10$. In case $T_{\mathrm{max}}\lesssim 10^{12}\GeV$ the produced $X$ is negligible, we shall see in next section that inflation could then play an important role.

\section{Effects of Inflation Dynamics}\label{sec:inflation}

\begin{table}[bt]
	\begin{tabular}{|c|c|c|c|c|}
		\hline 
		\diagbox{Initial $i(m)$}{Final $f(M)$} &  $S$  & $F$ & $V$ & $\gamma$	\\ [5pt]
		\hline 
		$S$ & $\dfrac{1}{32} \left[2 (1-6\zeta)m^2+M^2\right]^2 $ & $\dfrac{1}{16} M^2 \left(m^2-M^2\right) $ & $\dfrac{1}{32} \left(4 m^4-4 m^2 M^2+3 M^4\right)$ & $0$\\ [5pt]
		\hline
		$F$ & $0$ &$ 0 $ & $0 $ & $0$ \\ 
		\hline
		$V$ & $\dfrac{1}{32} \left(12 m^4-20 m^2 M^2+11 M^4\right)$ & $m^4-\dfrac{5 m^2 M^2}{16}-\dfrac{11 M^4}{16}$ & $\dfrac{1}{32} \left(140 m^4+148 m^2 M^2+33 M^4\right)$ & $4 m^4 $ \\ [5pt]
		\hline
		$\gamma$&$0$ & $0$ & $0$ & $0$ \\ 
		\hline 
	\end{tabular}
	\caption{$\mathcal{A}$ for the case $s=4m^2$(initial states with mass $m$ and final states with mass $M$). As usual, $m$ is the mass of initial particle and it is equal to zero for $\gamma$. For scalar as final state in the second row, we also include the non-mimimal coupling, $\zeta R S^\dagger S$.  \label{tab:As4m}}
\end{table}

It is widely believed that in the very early universe there was an exponential expansion called inflation. After inflation, there was a short matter-dominant time as the inflation field $\phi$ oscillates around the potential minimum. Then inflatons decay into radiation with decay width $\Gamma_{\phi}$ and reheat the universe with a temperature $T_R\sim \sqrt{\Gamma_{\phi}M_P}$. In the simplest approximation, we may just take $T_{\mathrm{max}}=T_R$ in our above discussion. However, realistically the effects from inflation are model-dependent since different inflationary scenarios could give various cosmological evolutions. More importantly, inflatons can also annihilate gravitationally into other particles and contribute the production. Here, we discuss some possible effects and only focus on the simplest chaotic inflation~\footnote{Dark matter produced by vacuum fluctuation in inflation with Coleman-Weinberg potential is recently studied in Ref.~\cite{Kannike:2016jfs}, whose starting point is different from ours.}, for example, with quadratic potential, although our formalism might also apply for other cases. 

For inflation field $\phi$ with canonical kinetic term and general potential $V(\phi)$, its energy-momentum tensor is given by
\begin{align}
T^{\mu\nu}_\phi=&-\eta^{\mu\nu}\left[\frac{1}{2}\partial ^\alpha \phi \partial _\alpha \phi - V(\phi)\right] + \partial^\mu \phi \partial ^\nu \phi.
\end{align}
During inflation, we can use homogeneous field configuration $\phi(t)$ for the background evolution and get the energy and pressure densities 
\begin{equation}
\rho_\phi = \frac{1}{2}\dot{\phi}^2 + V(\phi), p_\phi = \frac{1}{2}\dot{\phi}^2 - V(\phi).
\end{equation}
Using the equation of motion for $\phi$~\cite{Kolb:1990vq}
\begin{equation}
	\ddot{\phi} + 3H \dot{\phi} + \Gamma_{\phi} \dot{\phi} + V'(\phi)=0,
\end{equation}
where $\Gamma_{\phi}$ is the decay width that closely connects to the reheating that inflatons decay into other particles and reheat the Universe, we can obtain the evolution equation for $\rho_\phi$,
\begin{equation}
\dot{\rho}_\phi + 3H \dot{\phi}^2 = - \Gamma_{\phi}\dot{\phi}^2.
\end{equation}
Usually, averaging over several oscillations is performed so that one can use Virial theorem to replace $\dot{\phi}^2$ with averaged $\bar{\rho}_\phi$, and $\bar{\rho}_\phi$ just follows the evolution equation for non-relativistic matter. However, we do not perform such an average as we shall see immediately that the dominant production from inflaton happens at the transition time, not at the oscillation time. To compute the particle production from inflaton annihilation, we treat inflatons as particles with zero spatial momentum, namely the distribution of $\phi$ particle is
\begin{equation}
f=n_\phi (2\pi)^3 \delta^3(\bm{p}),\; n_{\phi}= \rho_\phi/m_\phi, \label{eq:dist}
\end{equation}
where $m_\phi$ is the mass of inflaton. We are aware that inflation field can not be always treated as collection of inflaton particles, for example, if some particle couples to inflaton non-gravitationally, its production should be calculated by solving the equation of motion and regarding inflation field as classical background~\cite{Kofman:1997yn}. Since in our case there is no direct coupling between other particles and inflaton, we simply use Eq.~(\ref{eq:dist}) for our estimations. This reduces Eq.~(\ref{eq:master}) to a very compact formula
\begin{align}
\frac{d\left(a^3n_X\right)}{a^3dt}= \frac{n^2_\phi }{m^2_\phi}\mathcal{F}\sigma = \frac{\rho^2_\phi }{m^4_\phi}\mathcal{F}\sigma . 
\end{align}
Note that generally $\mathcal{F}\sigma$ at $s=4m^2_\phi$ does not vanish because the factors $(s-4m^2_\phi)$ in both $\mathcal{F}$ and $|p_i|$ of $\sigma$ in Eq.~\ref{eq:cross} cancel with each other, unless there might be helicity/parity selection rules for different initial and final states so that the integrated squared matrix elements $\mathcal{A}$ is identically zero. We can see several examples in the second row of Table.~\ref{tab:As4m} with scalar as the initial state. For instance, for conformal coupled massless scalar ($\zeta=1/6$), massless fermions and vectors, $\mathcal{A}$ vanishes, which means that they are not produced by inflaton's gravitational annihilation during inflation.

For massive scalars and vectors, we have $\mathcal{F}\sigma\simeq \kappa^4 m^4_\phi/256\pi$ and can roughly estimate how much particles are produced at inflation during a Hubble time interval, $\Delta t \sim 1/H_\ast$,
\begin{equation}
n_X \simeq \frac{\rho_\phi^2 \kappa^4}{768\pi H_\ast}= \frac{3 H_\ast^3}{16\pi},
\end{equation}
where $H_\ast$ is the Hubble parameter around the transition era between inflation and oscillation time.
Interestingly, the above formula has the same power dependence on Hubble parameter as particle creation from vacuum fluctuations during inflation~\cite{Ford:1986sy, Shtanov:1994ce, Lyth:1996yj, Kofman:1997yn, Chung:1998ua, Kuzmin:1999zk} and oscillation~\cite{Ema:2015dka, Ema:2016hlw}. Moreover, the feature that conformal coupled massless scalars with $\zeta=1/6$, massless fermions and vectors are not produced by inflaton's gravitational annihilation during inflation also agrees with the results for vacuum fluctuations. This might be just a coincidence, or imply some deep underlying connection between two mechanisms, which however is beyond our scope here. In some sense the calculation with non-minimal coupling $\zeta$ serves as additional check of our computation.

\begin{figure}[t]
	\includegraphics[width=0.56\textwidth,height=0.46\textwidth]{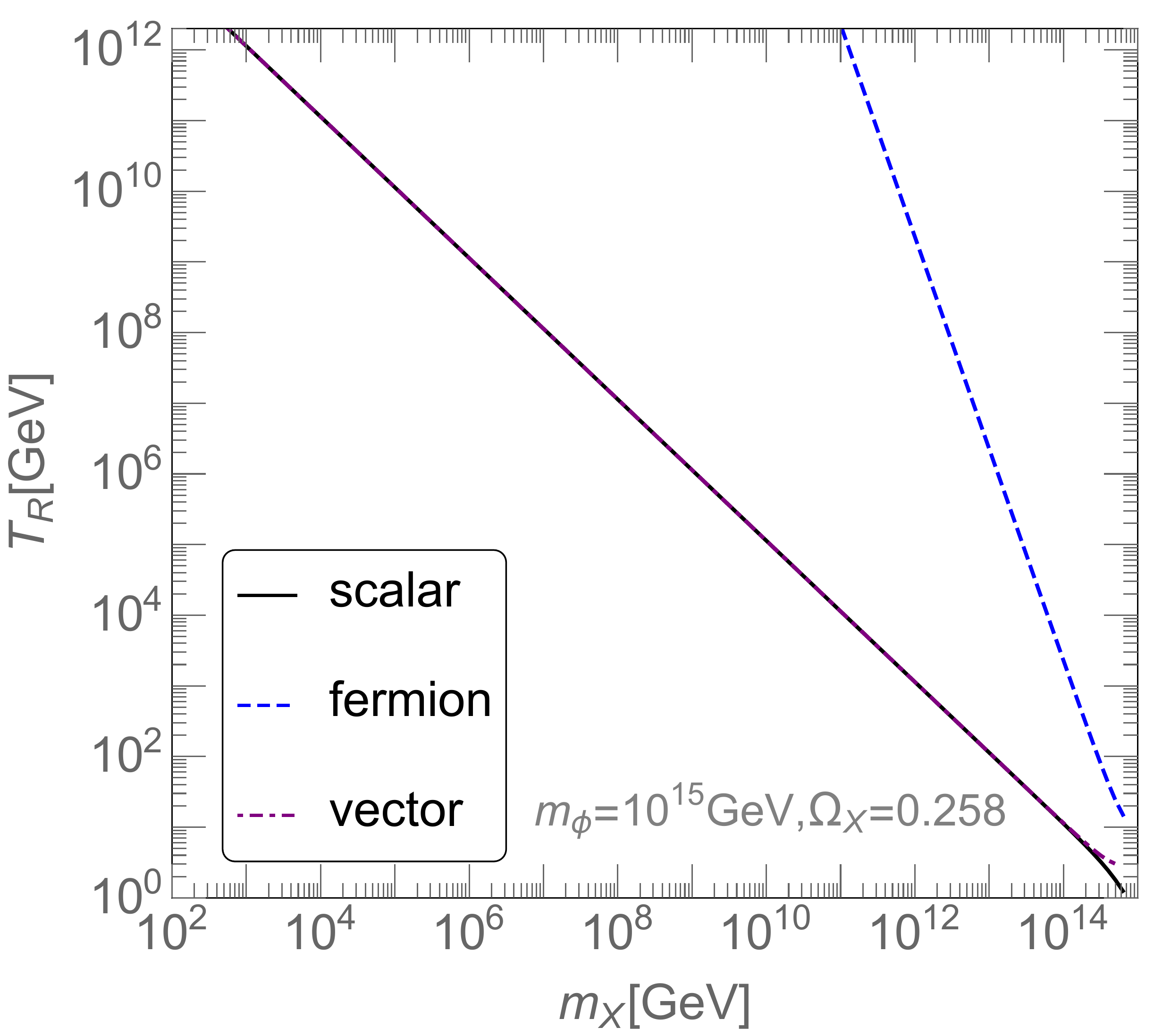}
	\caption{Contour $\Omega_X\simeq 0.258$ on the reheating temperature $T_R$ and DM mass $m_X$, in spin-0 (solid black line), spin-$\frac{1}{2}$ (blue dashed line) and spin-1 (purple dot-dashed) cases. We have used $m_\phi \simeq 10^{15}\GeV$ as example. Spin-0 coincides with spin-1 case except in the high mass regime when $m_X$ is close to $m_\phi$, which can be easily understood from $\mathcal{A}$ in Table.~\ref{tab:As4m}. \label{fig:relic2}}
\end{figure}

To get the yield, take $H_\ast=m_\phi$ and assume instantaneous reheating, we have $T_{\mathrm{max}}\simeq T_R = \sqrt{\Gamma_{\phi} M_P}$ and 
\begin{equation}
Y_X\simeq \frac{H_\ast}{M^2_{P}}T_R \simeq \frac{m_\phi}{M_{P}}\left(\frac{\Gamma_\phi}{M_{P}}\right)^{1/2}.
\end{equation}
From the above estimations, we can also learn that gravitational annihilation from inflatons might be dominant over the contributions from thermalized particles after reheating since the later one goes like $T^3_R/M_P^3$. 

From Fig.~\ref{fig:relic}, it is obvious that for $T_{\mathrm{max}}\lesssim 10^{12}\GeV$ the thermally produced $X$ is negligible. On the other hand, annihilation from inflations could still produce $X$ too abundantly for large $m_\phi$, unless $m_X$ and/or $\Gamma_{\phi}$ is small enough, for instance, $m_X \lesssim 1 \TeV$ for $\Gamma_{\phi}\sim 10^{-9} m_\phi\simeq 10^{-14}M_P$. However, for massive fermions there would be a suppression factor $m_X^2/m^2_\phi$ from the annihilation cross section. This feature is in sharp contrast with the contributions from thermalized particles discussed in previous section where production for particles with different spins are at the same order. In Fig.~\ref{fig:relic2}, we show how the contour $\Omega_X=0.258$ goes in the $m_X$-$T_R$ plane for spin-0, 1/2 and 1 with  $m_\phi=10^{15}\GeV$ and $T_R\lesssim 10^{12}\GeV$ as an example. One can easily see the dramatic difference discussed just above between spin-1/2 and spin-0/spin-1, which shows that generally spin-0/1 DM would require much lower reheating temperature or lighter mass. We also notice that spin-0 coincides with spin-1 case except in the high mass regime when $m_X$ is close to $m_\phi$, which can be easily understood from $\mathcal{A}$ in Table.~\ref{tab:As4m} because the longitude mode dominates in the high energy limit.

\section{Summary}\label{sec:con}
We have investigated the particle production from gravitational annihilation of thermal particles in the very early universe. In the case that dark matter (DM) particle might only have gravitational interaction, we have calculated the relic abundance and the possible viable mass range for DM with spin 0, 1/2 and 1. We have computed the analytical cross section for general gravitational annihilation processes through a graviton. DM could be produced by gravitational annihilation of all other particles in the thermal path after inflation or inflatons during inflation. The first contribution crucially depends on the highest temperature $T_{\mathrm{max}}$ after inflation, proportional to $T^3_{\mathrm{max}}/M^3_P$ if DM mass $m_X<T_{\mathrm{max}}$ and $m^3_X/M^3_P \exp{[-2m_X/T_{\mathrm{max}}]}$ if $m_X>T_{\mathrm{max}}$. Particles with different spins produced by thermal bath are at similar order and can give the correct abundance for DM with mass between $\TeV$ and $10^{16}\GeV$. While the second contribution from inflaton depends on the inflation scale, reheating temperature and also the spin of DM (spin 1/2 case is suppressed due to helicity selection, compared with scalar and vector particles). We have shown that in simplest chaotic inflation model the contribution from inflaton's gravitational annihilation could be the dominant production mechanism for stable particles like DM.

\acknowledgments

YT would like to thank Takeo Moroi for enlightening discussions and reading the manuscript, and is grateful to Kazunori Nakayama and Yohei Ema for helpful discussions. YT is supported by the Grant-in-Aid for Innovative Areas No.16H06490.

{\it Note Added:} While we were finalizing the manuscript, a preprint~\cite{Kolb:2017} appeared, which discussed the particle production from Higgs portal due to effective operators suppressed by Planck scale.

%\appendix

%\bibliographystyle{../JHEP}
%\bibliography{../references}

\providecommand{\href}[2]{#2}\begingroup\raggedright\endgroup

\end{document}